\documentclass[a4paper,linenumbers,11pt]{article}
\usepackage{pos}
\usepackage{siunitx}
\usepackage{multirow}
\usepackage{import}

\graphicspath{%
	{figures/}%
}

\import{./}{00_definitions}

\def\Journal#1#2#3#4{{#1} {\bf #2}, #3 (#4)}

\def\CPC{\em Comput. Phys. Commun.}
\def\CSBS{\em Comput. Softw. Big Sci.}
\def\EPJC{{\em Eur. Phys. J.} C}
\def\JHEP{\em J. High Energ. Phys.}

\def\NIMA{{\em Nucl. Instrum. Meth.} A}

\def\PLB{{\em Phys. Lett.}  B}
\def\PRL{\em Phys. Rev. Lett.}
\def\PRD{{\em Phys. Rev.} D}
\def\PTEP{\em Prog. Theor. Exp. Phys.}
\def\ZPC{{\em Z. Phys.} C}

\def\Vcb{\ensuremath{|V_{cb}|}\xspace}

\def\cosBY{\ensuremath{\cos\theta_{BY}}\xspace}
\def\Eecl{\ensuremath{E_\text{ECL}}\xspace}

\title{Lepton universality tests and searches for new physics in charged current decays at Belle~II}
\ShortTitle{Lepton universality tests in charged current decays at Belle II}

\manuallySeparateAuthors
\author*[a]{Henrik Junkerkalefeld}
\author{ on behalf of the Belle~II Collaboration}

\affiliation[a]{Physikalisches Institut, Universität Bonn,\\
	Nussallee 12, 53115 Bonn, Germany}

\emailAdd{junkerkalefeld@physik.uni-bonn.de}

\abstract{
	We present recent tests of lepton universality as crucial probes of the Standard Model in semileptonic \B-meson decays at Belle~II. All presented analyses use a data sample collected at the \FourS resonance by the Belle~II experiment corresponding to an integrated luminosity of \SI{189}{\per\femto\barn}. We report three analyses that probe light-lepton universality: The first measurement of a complete set of five angular asymmetries using $\Bzb \to \Dstp \ell^- \nulb$ decays, followed by a measurement of the forward-backward asymmetry in untagged $\Bzb \to \Dstp \ell^- \nulb$ decays, and finally, measurements of the branching-fraction ratios $\RDstemu = \BF(\Bzb \to \Dstp e^- \nueb) / \BF(\Bzb \to \Dstp \mu^- \numub)$, using the same untagged data set, along with the first measurement of the inclusive light-lepton ratio $\Remu = \BF(\bxenu)/\BF(\bxmunu)$.\\
	Furthermore, we present two tests of heavy-to-light lepton universality. In the first test, we report the measurement of the branching-fraction ratio \RDst using hadronic tagging. In the second test, we present the first measurement of the inclusive ratio \RX.\\
	All presented results are consistent with their corresponding Standard Model predictions and, where applicable, with the experimental world averages.
}

\FullConference{
16th International Conference on Heavy Quarks and Leptons (HQL2023)\\
28 November-2 December 2023\\
TIFR, Mumbai, Maharashtra, India\\}

\begin{document}
	\maketitle

\section{Introduction}
\label{sec:intro}

In the Standard Model of particle physics (SM) electroweak gauge bosons couple identically to the charged leptons $(e, \mu, \tau)$, a symmetry referred to as lepton universality (LU).
Semileptonic $B$-meson decays provide excellent sensitivity to potential lepton-universality violating (LUV) phenomena beyond the SM as high precision can be achieved both in experimental measurements and theoretical predictions. Due to the cancellation of systematic uncertainties associated with common factors, the precision is increased further when properties, like angular asymmetries or branching fractions, are compared directly between lepton flavors.\par

In a reinterpretation of Belle data~\cite{belle_remu}, $3.9\,\sigma$ evidence for LUV has been reported  in angular distributions of $\Bb \to D^* \ell^- \nulb$ decays in Ref.~\cite{bobeth}. Here and throughout, $\ell$ refers to a light lepton, specifically either an electron or a muon. In Sections \ref{sec:afbpp} to \ref{sec:remu}, we present three tests of light-lepton universality.\par
Additionally, the fundamental principle of LU is challenged in a combination of several measurements of the ratios $\RDorDst= \BF(\bdordsttaunu)/\BF(\bdordstlnu)$~\cite{babar_1, babar_2, belle_hadronic, belle_polarization_1, belle_polarization_2, belle_semileptonic, lhcb_1, lhcb_2}, resulting in a tension between experimental results and SM expectations of $3.3\,\sigma$~\cite{hflav}. In Sections \ref{sec:rdst} and \ref{sec:rx}, we present two results that test this tension.\par
Charge conjugation is implied in all physical processes and natural units are used.

\section{The \belletwo detector, experimental data and simulated samples}
\label{sec:detector_and_ds}

The Belle~II experiment is a general-purpose experiment situated at the SuperKEKB asymmetric-energy electron-positron collider~\cite{superkekb}. The accelerator delivers collisions at a center-of-mass energy of $\sqrt{s}=10.58$~GeV, corresponding to the mass of the \FourS resonance, which decays almost exclusively into a pair of  \B mesons.\par
The \belletwo detector~\cite{b2tdr, b2tip} encloses the interaction point of the collider within a cylindrical structure, comprising various subsystems. Positioned closest to the beam pipe is the vertex detector, equipped with several layers of silicon pixels and silicon strip detectors. The central drift chamber (CDC), filled with a gas mixture of $\mathrm{C}_2\mathrm{H}_6$ and He in equal parts, is used to measure particle trajectories. Encompassing the CDC are a Cherenkov-light imaging and time-of-propagation detector in the barrel region and a proximity-focusing, ring-imaging Cherenkov detector in the forward endcap region. These components furnish essential particle-identification details, especially facilitating the discrimination between kaons and pions. 
The electromagnetic calorimeter (ECL), comprising a barrel and two endcap regions, enables neutral particle and electron identification. A superconducting solenoid envelops the ECL, generating a uniform 1.5~T magnetic field along the beam direction.
The \belletwo configuration culminates with the \KL and muon identification detector, featuring scintillator strips in the inner part and endcaps of the barrel, accompanied by resistive plate chambers in the outer barrel.\par
In all the analyses presented, we rely on collision data collected from 2019 to 2021, corresponding to an integrated luminosity of \SI{189}{\per\femto\barn} at the \FourS resonance. Additionally, \SI{18}{\per\femto\barn} of data collected \SI{60}{\MeV} below the resonance is employed to determine continuum processes $e^+e^- \to \qqbar$.\par
To model signal and background processes, we use Monte Carlo simulation. For the simulation of $e^+e^- \to \FourS \to \BB$ processes and their subsequent decays, we utilize the software packages \evtgen~\cite{evtgen} and \pythia~\cite{pythia}. In the case of $e^+e^- \to \qqbar$, the software packages \kkmc~\cite{kkmc} and \pythia are used. To account for final-state radiation of photons from charged particles, we use \photos~\cite{photos}. The detector response and material interaction is simulated using \geant~\cite{geant}. 
The reconstruction and analysis of both simulated and experimental data is performed using the Belle II analysis software framework, \basftwo~\cite{basf2, basf2_repo}.

\section{Measurement of a complete set of angular asymmetries in $\Bzb \to \Dstp \ell^- \nulb$}
\label{sec:afbpp}
In Ref.~\cite{afbpp}, we present the first measurement of a complete set of angular asymmetry observables, called $A_\text{FB}$, $S_3$, $S_5$, $S_7$, and $S_9$, as a test of light-lepton universality. In a comparison of these observables between electrons and muons, most theoretical and experimental uncertainties cancel, resulting in high sensitivity to LUV~\cite{angular_asymm}. They are obtained by formulating disjoint one- or two-dimensional integrals of the differential decay rate as a function of the recoil parameter $w = (m_B^2+m_{D^*}^2-q^2)/(2 m_B m_{D^*})$ and the helicity angles $\cos\theta_\ell$, $\cos\theta_V$, and $\chi$. These angles correspond to the orientations between the lepton and $W$ boson directions, $D$ and $D^*$ meson directions, and the angle between the two decay planes, respectively. Expressed mathematically, the angular asymmetries take the form:
\begin{equation}
	\mathcal{A}_x(w) = \left( \frac{\text{d}\Gamma}{\text{d}w}\right)^{-1} \left[ \int_0^1 - \int_{-1}^0 \right]\text{d}x \frac{\text{d}^2\Gamma}{\text{d}w\text{d}x} \; \text{,}
\end{equation}
where $x = \cos\theta_\ell$ for $A_\text{FB}$, $\cos 2\chi$ for $S_3$, $\cos\chi\cos\theta_V$ for $S_5$, $\sin\chi\cos\theta_V$ for $S_7$, and $\sin 2\chi$ for $S_9$. The test of light-lepton universality involves comparing the angular asymmetries of electrons and muons, denoted as $\Delta \mathcal{A}_x (w) = \mathcal{A}_x^\mu(w) - \mathcal{A}_x^e(w)$. Measurements are conducted in three distinct ranges of $w$: The full phase space ($w_\text{incl.}$), $w \in [1.0, 1.275]$ ($w_\text{low}$) and $w \in [1.275, \approx 1.5]$ ($w_\text{high}$).\par
We use the Full Event Interpretation (FEI) algorithm to reconstruct one of the \B mesons, the \btag, in a fully hadronic decay mode~\cite{FEI}. In events featuring such a candidate, we proceed to reconstruct $\Bzb \to \Dstp \ell^- \nulb$ candidates with $\Dstp \to \Dz \pi^+$ and $\Dz \to K^-\pi^+$, $K^-\pi^+\pi^0$, $K^-\pi^+\pi^+\pi^-$, $K^-\pi^+\pi^+\pi^-\pi^0$, $K_\text{S}^0\pi^+\pi^-$, $K_\text{S}^0\pi^+\pi^-\pi^0$, $K_\text{S}^0\pi^0$, or $K^-K^+$. The selection criteria ensure that all tracks originate from the $e^+e^-$ interaction point, and lepton candidates are identified with likelihood ratios based on the different subdetectors.\par
The signal and background yields are determined through binned maximum-likelihood fits to distributions of $\mmsq = ((\sqrt{s}, \vec{0}) - P^{\,\cms}_{\btag} - P^{\,\cms}_{\Dst} - P^{\,\cms}_\ell)^2$. Here $P^{\,\cms}_{\btag}$, $P^{\,\cms}_{\Dst}$, and $P^{\,\cms}_\ell$ are the measured center-of-mass-frame four-momenta of the \btag candidate, \Dst meson, and the signal lepton, respectively. These distributions exhibit a peak at $\mmsq \approx \SI{0}{\GeV^2}$ for correctly reconstructed signal events and are more widely distributed for background events. The fitted yields are corrected  for selection efficiency and detector acceptance, as well as for bin migrations. Statistical uncertainties limit the measurement, typically being one order of magnitude larger than uncertainties arising from bin migrations or systematic uncertainties.\par

\begin{figure}[htbp]
	\centerline{\includegraphics[width=\linewidth]{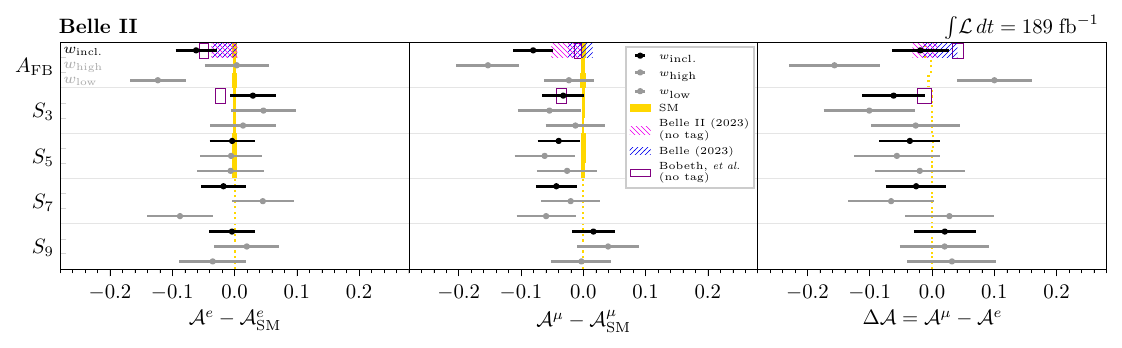}}
	\caption{The measured asymmetries and asymmetry differences are illustrated, allowing for a comparison with recent results from Belle~\cite{mprim} and Belle~II (cf. Sec. \ref{sec:untaggedDst}). Additionally, the presentation includes calculations from Bobeth \textit{et al.} \cite{bobeth} and the theory expectations. Figure from Ref.~\cite{afbpp}.}
	\label{fig:afb_results}
\end{figure}

The findings are presented in Figure~\ref{fig:afb_results}. The $\chi^2$ tests conducted in each of the three $w$ regions show good agreement with expectations from the SM. The minimum $p$ value obtained is $0.13$, indicating no significant departure from expectations and, consequently, no evidence for LUV is observed.
	
\section{Light-lepton universality tests in untagged $\Bzb \to \Dstp \ell^- \nulb$}
\label{sec:untaggedDst}

In the analysis documented in Ref.~\cite{untaggedDst}, we reconstruct $\Bzb \to \Dstp \ell^- \nulb$ decays. The central goal of this analysis is to measure the CKM matrix element \Vcb and the branching fraction $\BF(\Bzb \to \Dstp \ell^- \nulb)$. Additionally, the analysis explores several observables that are particularly well-suited for investigating light-lepton universality.\par
In this analysis, the identification of light leptons is accomplished through particle identification likelihood ratios, and the \Dst meson is reconstructed through the decay chain $\Dstp \to \Dz \pi^+$ with subsequent $\Dz \to K^-\pi^+$ decays, ensuring appropriate invariant masses.\par
The accompanying $B$ meson originating from the \FourS decay is not explicitly reconstructed. Consequently, the momentum direction of the signal $B$ meson is estimated through a novel approach. Assuming correct reconstruction, the angle \cosBY between the signal $B$ meson and the $Y = D^* + \ell$ system is deduced from beam energy information and the kinematics of the reconstructed particles. A weighted average of potential momentum directions along the cone created by \cosBY is computed. 
The weights reflect the prior probability of the anticipated angular distribution of $B$ mesons along the beam axis due to the $\Upsilon(4S)$ polarization and additionally consider rest-of-event data from tracks and clusters not associated with the $Y$ system.\par
The derived directional information serves as the basis for determining the recoil parameter $w$ and the helicity angles $\cos\theta_\ell$, $\cos\theta_V$, and $\chi$, as introduced in Section \ref{sec:afbpp}. The four parameters are divided into 10 (8) equidistant intervals each. Background and signal yields are individually obtained through a two-dimensional maximum-likelihood fit to \cosBY and the mass difference between the \Dstp and \Dz mesons, denoted as $\Delta M$. Subsequently, corrections are applied to account for the effects of finite reconstruction resolution, efficiency, and detector acceptance.\par

We obtain the angular asymmetry $A_\text{FB}$ from the $\cos\theta_\ell$ distribution for each lepton flavor and find
\begin{align}
	A_\text{FB}^e &= 0.228 \pm 0.012 \pm 0.018 \text{,} \\
	A_\text{FB}^\mu &= 0.211 \pm 0.011 \pm 0.021 \text{, and} \\
	\Delta A_\text{FB} &= (-17 \pm 16 \pm 16)\times 10^{-3} \text{.}
\end{align}
Here and throughout, the first contribution to the uncertainty is statistical and the second is systematic. In Figure~\ref{fig:afb_results}, the result is illustrated in the pink hatched area. By summing the partial decay rates of all kinematic variables and averaging over $w$ and the helicity angles, we derive the branching fractions for each lepton flavor and their ratio 
\begin{equation}
	\RDstemu = \BF(\Bzb \to \Dstp e^- \nueb) / \BF(\Bzb \to \Dstp \mu^- \numub) = 0.998 \pm 0.009 \pm 0.020      \text{.}
\end{equation}
This result aligns with SM predictions, which deviate slightly from each other, specifically $1.0026 \pm 0.0001$~\cite{bobeth} and $1.0041 \pm 0.0001$~\cite{bernlochner_rdstemu}. Additionally, our findings are compatible with experimental results of the same ratio, as reported in Refs.~\cite{mprim, belle_remu}.

\section{Measurement of the inclusive ratio \Remu}
\label{sec:remu}

In this analysis, published in Ref.~\cite{RXemu}, we report the first measurement of the inclusive ratio of branching fractions $\Remu = \BF(\bxenu) / \BF(\bxmunu)$.\par The  \btag meson is reconstructed in fully hadronic decays using the FEI algorithm. We require a lepton candidate, identified using a multiclass BDT~\cite{eBDT} for electrons and a likelihood ratio for muons. The lepton's charge must be consistent with the appropriate semileptonic decay of the accompanying \B meson, the signal \bsig, assuming opposite flavor of the \btag: $\bsig^{0, +} \to X \ell^{+} \nul$. Additionally, we impose a criterion that the lepton candidate must have a momentum in the rest frame of the \bsig meson $\plB > \SI{1.3}{\GeV}$ to mitigate contributions from hadrons misidentified as leptons, leptons emerging from secondary cascade decays, and lepton daughters of $\bxtaunu$ decays. To suppress continuum backgrounds, we employ a BDT trained on event-shape quantities. Any remaining continuum events are modeled using the off-resonance data sample.\par
We extract the signal yields in a binned maximum-likelihood fit to the \plB distributions of each lepton flavor. The fit is conducted simultaneously in the $e$ and $\mu$ channels, allowing for the consideration of correlated systematic uncertainties between both flavors, so that branching-fraction and form-factor uncertainties from \bxlnu decays mostly cancel in the ratio. Gaussian constraints are applied to \BB background yields using a fit to data in a background-enriched control channel. This control channel consists of events with two \B mesons reconstructed with the same flavor. To constrain continuum background yields, we utilize the off-resonance data.\par
We find
\begin{equation}
	\Remu = 1.007 \pm 0.09 \pm 0.019      \text{.}
\end{equation}
The primary contributors to the systematic uncertainties stem from uncertainties associated with the lepton-identification efficiency. Our result aligns well with the SM expectation of $1.006 \pm 0.001$~\cite{kvos}. This outcome represents the most precise test of electron-muon universality in semileptonic $B$ decays based on branching fractions to date.

\section{Hadronically tagged \RDst}
\label{sec:rdst}

We report Belle II's first measurement of the heavy-to-light-lepton branching-fraction ratio $\RDst = \BFdsttaunu / \BFdstlnu$. The reconstruction of one \B meson is achieved using the FEI algorithm with hadronic tagging. On the signal side, the $\tau$ lepton in the signal decay \bdsttaunu is required to decay leptonically, i.e., \tauenunu or \taumununu, so that the visible final-state particles align with the normalization mode \bdstlnu. Leptons are identified based on likelihood ratios. The \Dst meson is reconstructed in the decays $\Dstp \to \Dz \pip$, $\Dstp \to \Dp \piz$, and $\Dstz \to \Dz \piz$, with appropriate invariant-mass and vertex requirements. The \D mesons are reconstructed in the decay channels $\Dz \to K^-\pip$, $K^- \pip \piz$, $K^-\pip\pip\pim$, $\KS \pip \pim$, $\KS \pip \pim \piz$,  $\KS \piz$, $K^-K^+$, and $\pim \pip$, as well as $\Dp \to \KS \pip$, $K^-K^+ \pip$, and $K^-\pip \pip$.\par
The reconstruction of the full event allows us to impose a completeness constraint, specifically the requirement of no additional charged particle tracks in the event, to suppress misreconstructed backgrounds. We extract the signal in a two-dimensional maximum-likelihood fit in the residual energy in the ECL, denoted as \Eecl, and the missing mass squared \mmsq introduced in Section~\ref{sec:afbpp}. In correctly reconstructed signal and normalization processes, small \Eecl values are expected, while backgrounds feature larger values. Due to the three neutrinos in the final state, signal events are distinguishable from normalization and background events due to higher \mmsq values.\par
The modeling of \bdststlnu decays and non-resonant \bdordstpipilnu and \bdordstetalnu decays with experimentally insufficient constraints is validated using a $\Bb \to \Dst \piz \ell^- \nulb$ control channel. Backgrounds arising from misreconstructed \D mesons are calibrated in three different regions of \mmsq using a sideband region defined by the mass difference of \Dst and \D mesons $\Delta M$. We validate the modeling of \Eecl in a control region requiring $\mmsq < \SI{1}{\GeV^2}$ and find an excess of data at low values. To address this mismodeling, we subtract $(15 \pm 7)\,\text{MeV}$ from the energy of each neutral cluster originating from interactions of charged particles with detector material. The energy shift is determined by optimizing the $\chi^2$ agreement between experimental data and simulation.\par
In the signal extraction fit, the signal, normalization, and \bdststlnu yields are allowed to float freely. Backgrounds stemming from misreconstructed \D mesons are constrained based on their calibration in the $\Delta M$ control region, while remaining backgrounds are fixed according to their predicted branching fractions.\par
We find
\begin{equation}
	\RDst = 0.267\,^{+0.041}_{-0.039}\,^{+0.028}_{-0.033}  \text{.}
\end{equation}
The dominating sources of systematic uncertainty are the limited size of the simulation sample, the shapes of the \Eecl PDF, and the modeling of \bdststlnu decays. Our result, illustrated in Figure~\ref{fig:rx_and_rdst}, is consistent with both the SM prediction of $0.254 \pm 0.005$ and the current experimental world average $0.284 \pm 0.013$ as reported in Ref.~\cite{hflav}.

\section{First measurement of \RX}
\label{sec:rx}

A powerful test regarding the persistent tension in experimental results of \RDorDst and their SM predictions (cf. Sec.~\ref{sec:intro}), that is distinct in both its sensitivity to statistical and systematic uncertainties, is the measurement of the inclusive branching-fraction ratio $\RX = \BFxtaunu / \BFxlnu$. The LEP experiments measured a related branching fraction based on a $b$-hadron admixture in $\Z\to\bbbar$ processes, $\BF(b\text{-admix} \to \xtaunu)$, to be consistent with SM expectations~\cite{aleph, delphi, l3_1, l3_2, opal}. In Ref.~\cite{RX}, we present the first measurement of the inclusive ratio at the \FourS resonance, i.e., specifically focusing on \B mesons.\par
We reconstruct the \btag candidate in hadronic decays and require leptonic $\tau$-lepton decays, so that the signal process \bxtaunu and the normalization process \bxlnu share identical final-state particles in the reconstruction. Leptons are identified based on the multiclass BDT for electrons and likelihood ratios for muons. To mitigate background contributions from hadrons misidentified as leptons (fakes) and leptons from secondary decays (secondaries), we employ stringent lepton identification classifier thresholds. Moreover, muon fake contributions from pions or kaons are suppressed by rejecting combinations of the signal lepton candidate with other hadrons when invariant masses and vertices are consistent with specific $\omega$, $\Kstarz$, \Dz, or \Dp decays. Secondary leptons arising from photon conversion, $\piz \to e^+e^- (\gamma)$ decays, or $\jpsi \to \ell^+\ell^-$ are rejected if suitable combinations are identified in the event.\par
All tracks and neutral ECL clusters not associated with the \btag candidate or the signal lepton constitute the hadronic system $X$. Its kinematic properties are utilized to define \mx as the invariant mass of the $X$ system and to calculate quantities such as $\mmsq = ((\sqrt{s}, \vec{0}) - P^{\,\cms}_{\btag} - P^{\,\cms}_{X} - P^{\,\cms}_\ell)^2$, where $P^{\,\cms}_{i}$ denote the measured center-of-mass-frame four-momenta of the respective particle $i$.

\begin{figure}[htbp]
	\centering
	\sbox0{\includegraphics[width=\linewidth]{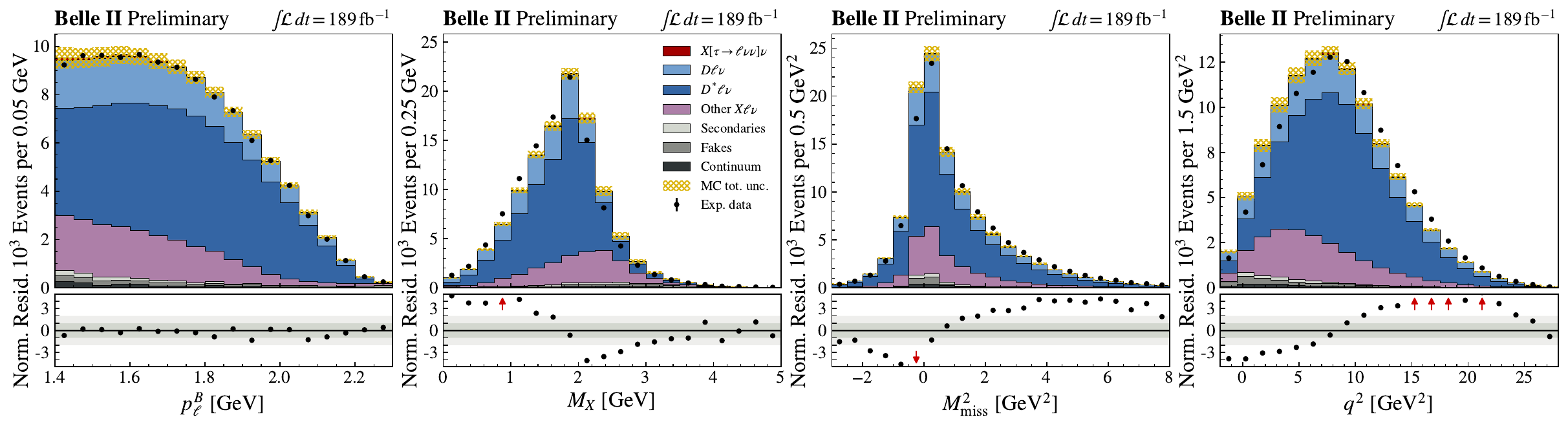}}
	\includegraphics[clip,trim={0.69\wd0} 0 {1.36\wd0} 0, width=0.32\linewidth]{plB_mx_mm2_q2_syst_highPLB_BB_ell_prescaled_publ_version_wFakes.pdf}
	\includegraphics[clip,trim=0 0 {1.36\wd0} 0, width=0.64\linewidth]{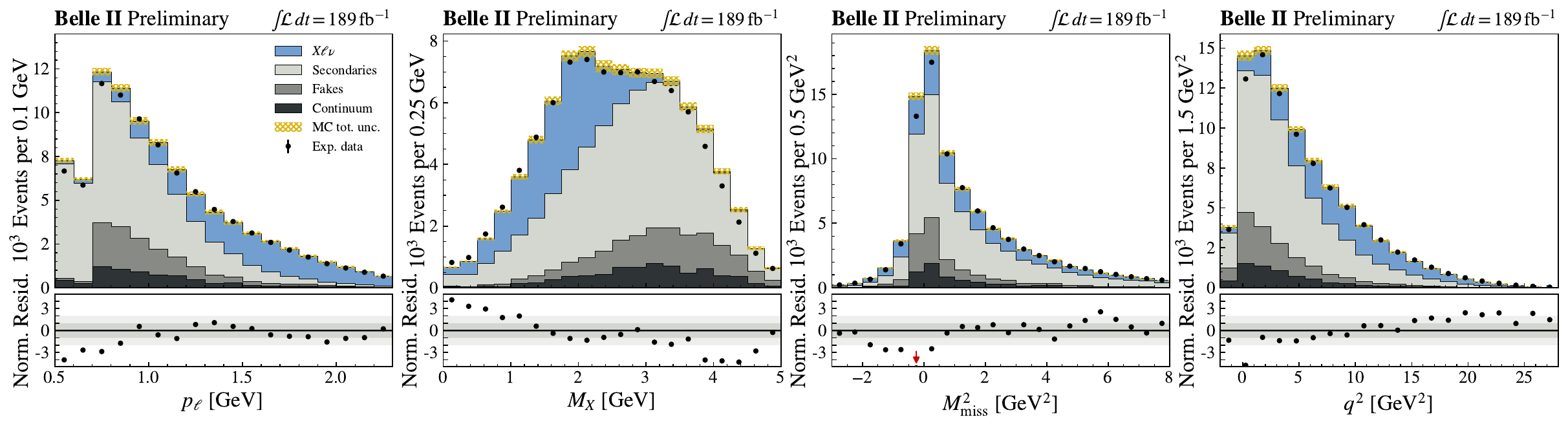}\\
	\includegraphics[clip,trim={0.69\wd0} 0 {1.36\wd0} 0, width=0.32\linewidth]{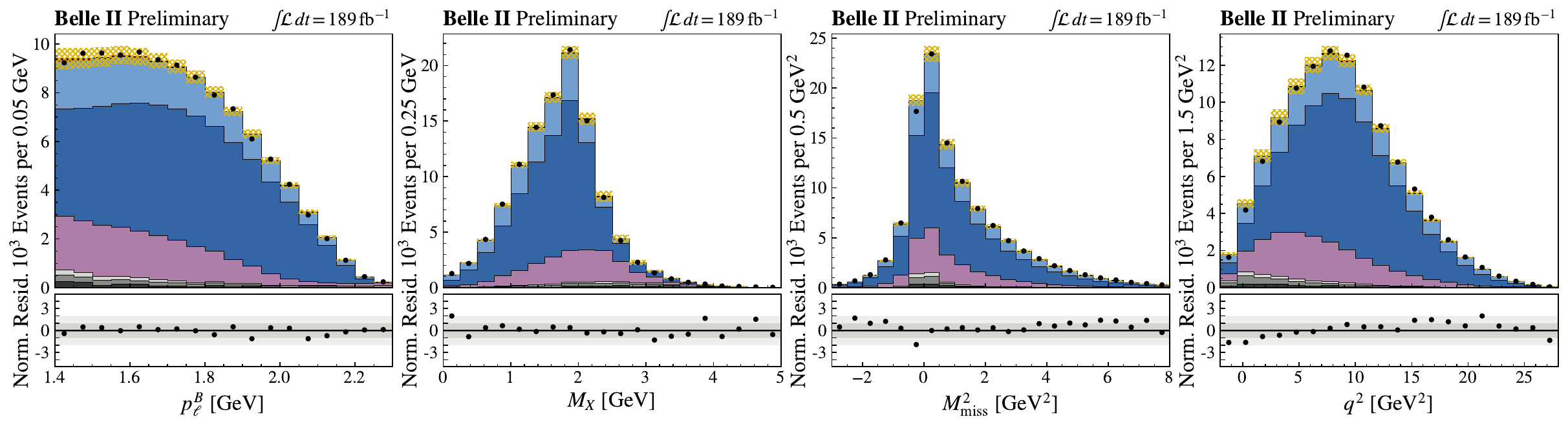}
	\includegraphics[clip,trim=0 0 {1.36\wd0} 0, width=0.64\linewidth]{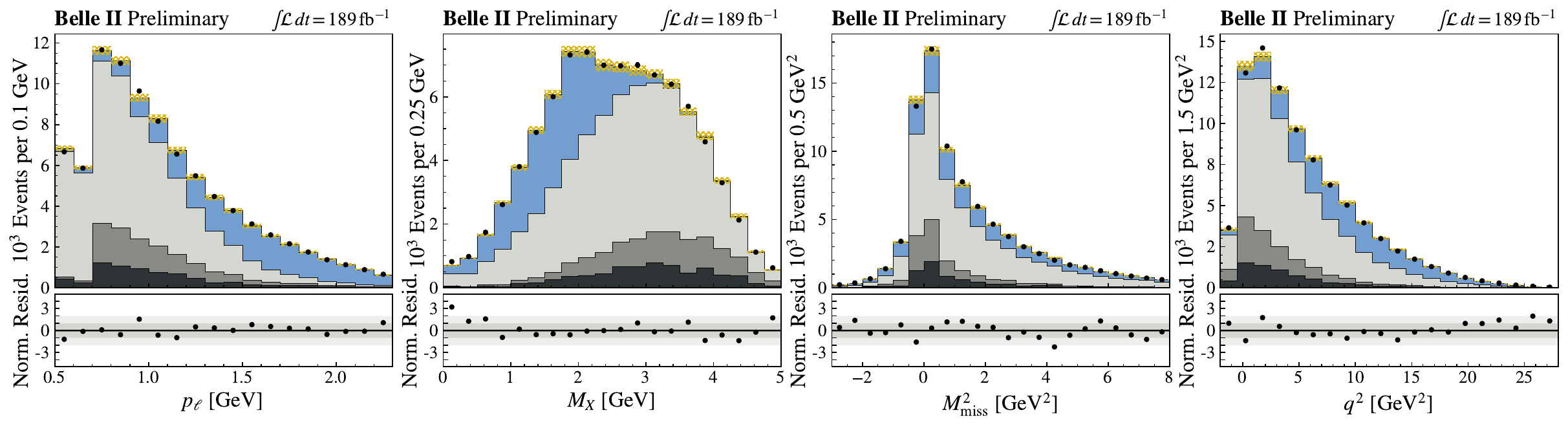}
	\caption{The distributions of quantities used in the reweighting procedure are presented. Normalization \bxlnu and signal \bxtaunu processes are reweighted based on the \mx distribution in the high \plB control region (left plot), while \BB background processes, i.e., fakes and secondaries, are reweighted in two-dimensional intervals of \pllab and \mx in the same-flavor control region (center and right plot). The top row of plots depicts the pre-reweighting distributions in simulated (filled histograms) and experimental (black points) data with their uncertainty-normalized disagreements shown below. The bottom row presents the reweighted distributions. }
	\label{fig:rx_reweighting}
\end{figure}

In Figure~\ref{fig:rx_reweighting}, we compare experimental data with simulated \mx distributions in the high-\plB control region $\plB > \SI{1.4}{\GeV}$, primarily composed of \bxlnu processes ($95\%$). In comparison to the simulation, a significant excess of experimental data is observed in the low-\mx region, while the high-\mx regions exhibits a deficit of data. The discrepancy between data and simulation suggests inappropriate modeling of \D decays, particularly those into \KL mesons. Since the corresponding inclusive \D-meson branching fractions are not measured with sufficient precision~\cite{pdg}, we introduce a data-driven reweighting approach. In this approach, \bxlnu and \bxtaunu processes are reweighted using the experimental-to-simulated yield ratio in intervals of \mx in the high-\plB control region. Secondary and fake processes are reweighted in two-dimensional intervals of \pllab and \mx using the same-flavor control region, where both \B mesons are reconstructed to have the same flavor. This results in a composition of $77\%$ fake, secondary, and continuum processes and $23\%$ \bxlnu decays from neutral \B-meson oscillations.\par
The reweighting procedure, based on distributions depicted in Figure~\ref{fig:rx_reweighting}, effectively addresses mismodeling in additional kinematic variables correlated to the $X$ system, such as \mmsq and \qsq, as detailed in the supplemental material of Ref.~\cite{RX}. This enables the extraction of the signal in a two-dimensional binned maximum-likelihood fit in \plB and \mmsq. In the fit, signal, normalization, and \BB background template yields float freely, while the continuum yields are constrained using off-resonance data. Both the electron and muon channels are simultaneously fit to correctly account for their correlation.\par
The results are
\begin{align}
	\RXe &= 0.232 \pm 0.020 \pm 0.037, \text{ and} \\
	\RXmu &= 0.222 \pm 0.027 \pm 0.050, 
\end{align}
respectively. A weighted average of correlated values yields
\begin{equation}
	\RX = 0.228 \pm 0.016 \pm 0.036      \text{.}
\end{equation}

\begin{figure}[htbp]
	\centerline{\includegraphics[width=0.7\linewidth]{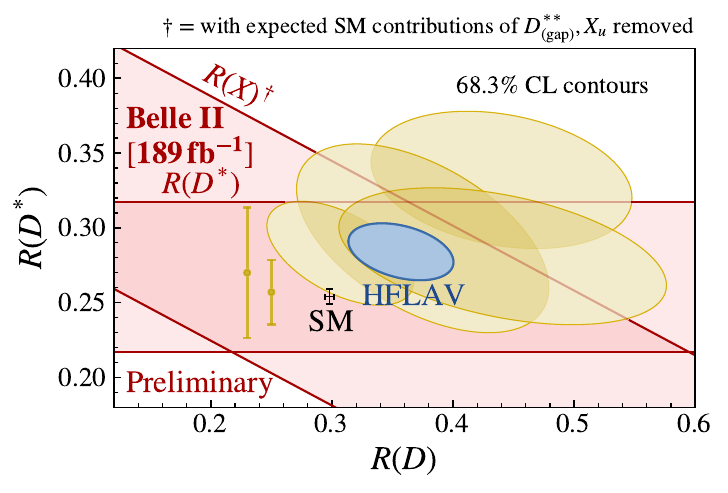}}
	\caption{Our presented \RDst and \RX~\cite{RX} results (red) compared to related individual measurements (gold)~\cite{babar_1, babar_2, belle_hadronic, belle_polarization_1, belle_polarization_2, belle_semileptonic, lhcb_1, lhcb_2} and the current world average (blue) as reported in Ref.~\cite{hflav}.}
	\label{fig:rx_and_rdst}
\end{figure}

Our result aligns with its SM prediction of $\RX_\text{SM} = 0.223 \pm 0.005$, as derived in an arithmetic average of Refs.~\cite{kvos, ligeti_rxc, ligeti_rxu}, and also is consistent with the current experimental world average of \RDorDst. The leading systematic uncertainties arise from the limited size of the simulation sample, the discrepancy (gap) in branching fractions between the inclusive measurement of \BFxclnu and the sum of exclusive measurements, variations in different form-factor model predictions for \bdstlnu process kinematics, and the statistical precision of the \mx-based reweighting procedures. We expect several of the major systematic uncertainties to decrease with a larger experimental data sample size like statistical uncertainties, implying that this measurement is statistically limited.\par
In Figure~\ref{fig:rx_and_rdst}, the \RX measurement is compared with experimental measurements and theoretical predictions of \RDorDst after subtracting the expected contributions of \bdststtaunu, $\Bb \to \gaptaunu$, and \bxutaunu from the measured value of \RX. However, more precise comparisons become feasible when the result is not constrained to the \RD-\RDst plane.

\end{document}